\documentstyle[12pt,epsf,rotating]{article}

\input axodraw.sty

\begin{document}

\catcode`@=11 \def\citer{\@ifnextchar
  [{\@tempswatrue\@citexr}{\@tempswafalse\@citexr[]}}

\def\@citexr[#1]#2{\if@filesw\immediate\write\@auxout{\string\citation{#2}}\fi
  \def\@citea{}\@cite{\@for\@citeb:=#2\do
    {\@citea\def\@citea{--\penalty\@m}\@ifundefined {b@\@citeb}{{\bf
          ?}\@warning
       {Citation `\@citeb' on page \thepage \space undefined}}%
\hbox{\csname b@\@citeb\endcsname}}}{#1}}
\catcode`@=12

\marginparwidth 3cm
\setlength{\hoffset}{-1cm}
\newcommand{\mpar}[1]{{\marginpar{\hbadness10000%
                      \sloppy\hfuzz10pt\boldmath\bf\footnotesize#1}}%
                  \typeout{marginpar: #1}\ignorespaces}
\def\mda{\mpar{\hfil$\downarrow$\hfil}\ignorespaces}
\def\mua{\mpar{\hfil$\uparrow$\hfil}\ignorespaces}
\def\mla{\marginpar[\boldmath\hfil$\rightarrow$\hfil]%
                   {\boldmath\hfil$\leftarrow $\hfil}%
                   \typeout{marginpar:
                     $\leftrightarrow$}\ignorespaces}

\providecommand{\LQ}{L\!Q}
\providecommand{\squ}{\tilde q}

\renewcommand{\abstractname}{Abstract}
\renewcommand{\figurename}{Figure}
\renewcommand{\thefootnote}{\fnsymbol{footnote}}

\begin{titlepage}

\begin{flushright}
DESY 97-43 \\
BI-TP 97/09 \\
March 1997 \\
\end{flushright}

\vspace{1cm}

\begin{center}
\baselineskip25pt

\def\thefootnote{\fnsymbol{footnote}}
{\large\sc Formation and Decay of scalar leptoquarks/squarks \\
in $ep$ collisions}

\end{center}

\setcounter{footnote}{3}

\vspace{1cm}

\begin{center}
\baselineskip12pt

{\sc 
T.~Plehn$^1$,
H.~Spiesberger$^{2,{\displaystyle \ast}}$,
M.~Spira$^3$,
and P.~M.~Zerwas$^1$} \\ 
\vspace{1cm}

$^1$ Deutsches Elektronen--Synchrotron DESY, D--22603 Hamburg

\vspace{0.3cm}

$^2$ Fakult\"at f\"ur Physik, Universit\"at Bielefeld, D--33501
Bielefeld 

\vspace{0.3cm}

$^3$ CERN, Theory Division, CH--1211 Geneva 23

\end{center}

\vspace{2cm}
\begin{abstract}
  \normalsize \noindent 
The cross sections for the formation of scalar resonances,
leptoquarks or squarks, in electron/positron--proton collisions
at HERA are presented including next-to-leading order 
QCD corrections. Depending mildly on the mass of 
the resonances, the $K$-factors increase the production cross sections
by up to 30\% if the target quarks are valence quarks. The QCD
corrections to the partial decay widths of leptoquarks/squarks to
leptons and quarks are small. The electron spectrum in the decays is
softened nevertheless by perturbative gluon radiation at a level of
3.4 GeV for a leptoquark/squark mass of 200 GeV.
\end{abstract}

\vspace*{\fill} 
\footnoterule 
{\footnotesize
  \noindent ${}^{\displaystyle \ast}$ Supported by Bundesministerium
  f\"ur Bildung, Wissenschaft, Forschung und Technologie, Bonn,
  Germany, Contract 05 7BI92P (9).}

\end{titlepage}

\def\thefootnote{\arabic{footnote}} \setcounter{footnote}{0}

\setcounter{page}{2}

\section{Introduction}
An excess of events has recently been observed in
deep-inelastic $e^+ p$ scattering at HERA for high $x$ and high $Q^2$
\cite{H1,ZEUS}.  If these events are attributed to the formation of a
resonance in $e^+ q$ collisions, decaying subsequently into
lepton--quark final states, the events should cluster at the invariant
mass $M=\sqrt{xs}$. Moreover, if scalar resonances are formed, the
distribution in the scaling variable $y$ should be flat, corresponding
to isotropic angular decay distributions in the rest frame of the
resonance. Indeed, a clustering around $M=200$~GeV is indicated in the
H1 data and the events are distributed up to large $y$ values
\cite{H1}. These observations have initiated theoretical analyses of
possible resonance formation at HERA, and cross checks with
experimental results from LEP \cite{LEP}, the Tevatron \cite{TEV}
and from rare decays have been performed. The analyses have been
carried out for leptoquarks in general and squarks in supersymmetric
theories with $R$-parity breaking in particular
\citer{alta,drei}.\bigskip

In the present report we present the cross sections for the formation
of scalar resonances, leptoquarks or squarks, including
next-to-leading order QCD corrections in $e^\pm p$ collisions.  The
step beyond the leading order analysis \cite{buch} is motivated by two
points.  First, the theoretical predictions for the cross sections are
stabilized with regard to variations of the renormalization and
factorization scales. The choice of the leptoquark/squark mass $M$,
which can {\it a priori} be considered as a natural choice for these
spurious parameters, can thus be investigated in detail. Second, to
derive the theoretical predictions for the formation cross sections to
an accuracy of ${\cal O}(\alpha_S)\sim 10\%$ is a reasonable goal for
the time being. The results are presented for a single scalar
resonance
\begin{equation}
e + q \rightarrow \LQ/\squ
\end{equation}
where $e$ generically denotes electrons and positrons, $q$ quarks and
antiquarks, and $\LQ/\squ$ leptoquarks and squarks (Fig.~\ref{fig1}a). The
coupling $\lambda$ is defined by
\begin{equation}
{\cal L}_{int} = \lambda (\bar{e} \; q) S + h.c.
\end{equation}
with the (opposite) helicities of the lepton and the quark fixed for the
specific leptoquark/squark state $S$ (see e.g. Table~1 of
Ref.~\cite{kali}).\bigskip

By the same token, we have determined the partial width of
leptoquark/squark decays
\begin{equation}
\LQ/\squ \rightarrow e + q (g)
\label{decay}
\end{equation}
to order $\alpha_s$ in QCD. For massless quarks in the final state, the
(scaled) lepton energy $x_e = 2E_e/M$ is fixed to be 1 in lowest
order. The lepton energy is softened by gluon radiation and the
spectrum becomes continuous, peaking however strongly for $x_e$ just
below 1.\bigskip

\section{Production of Scalar Particles in $ep$ Collisions}
The {\it cross section} in leading order QCD can be cast into the form
\begin{eqnarray}
\sigma_{LO} &=& \sigma_0 \; x \; q\left(x,M^2\right) \nonumber \\
\sigma_0    &=& \frac{\pi \lambda^2}{4 M^2}
\label{eq5}
\end{eqnarray}
with 
\begin{equation}
x = M^2/s
\end{equation}
The density of the target quarks at the scale $\mu_F = M$ is denoted
by $q(x,M^2)$.

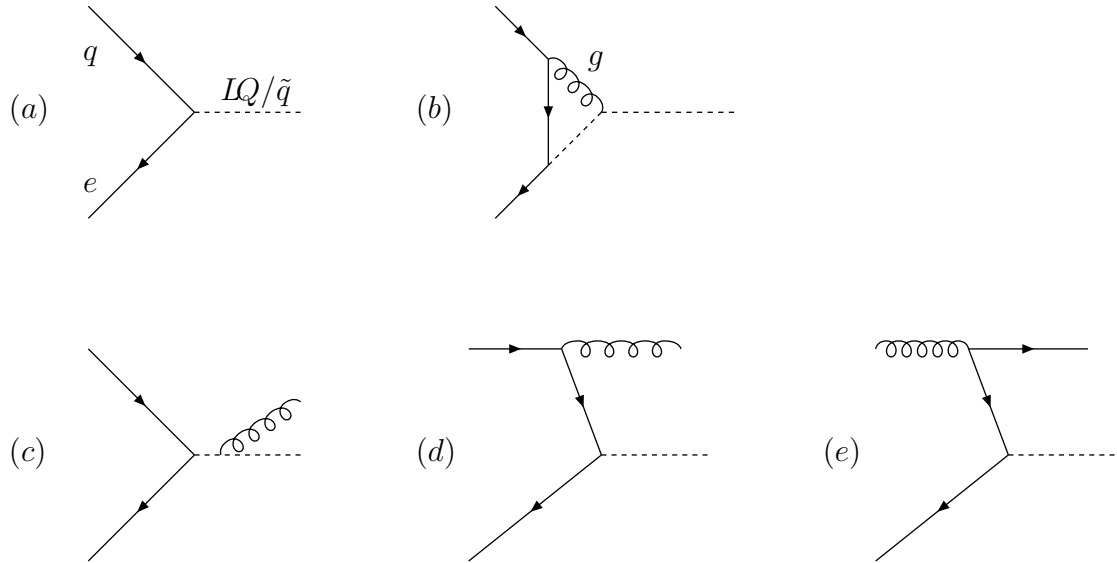
\begin{figure}[htb]

\begin{picture}(100,100)(-40,0)
\ArrowLine(10,90)(50,50)
\ArrowLine(50,50)(10,10)
\DashLine(50,50)(90,50){2}
\put(8,20){$e$}
\put(8,70){$q$}
\put(60,55){$\LQ/\squ$}
\put(-20,48){$(a)$}
\end{picture}
\begin{picture}(100,100)(-90,0)
\ArrowLine(10,90)(30,70)
\ArrowLine(30,70)(30,30)
\ArrowLine(30,30)(10,10)
\Gluon(30,70)(50,50){3}{3}
\DashLine(50,50)(30,30){2}
\DashLine(50,50)(100,50){2}
\put(45,68){$g$}
\put(-20,48){$(b)$}
\end{picture}  \\[0.5cm]

\begin{picture}(100,100)(-40,0)
\ArrowLine(10,90)(50,50)
\ArrowLine(50,50)(10,10)
\Gluon(60,50)(90,70){3}{4}
\DashLine(50,50)(90,50){2}
\put(-20,48){$(c)$}
\end{picture}
\begin{picture}(100,100)(-90,0)
\ArrowLine(0,90)(35,90)
\ArrowLine(50,50)(0,10)
\ArrowLine(35,90)(50,50)
\Gluon(35,90)(80,90){3}{4}
\DashLine(50,50)(90,50){2}
\put(-20,48){$(d)$}
\end{picture}
\begin{picture}(100,100)(-140,0)
\Gluon(0,90)(35,90){3}{5}
\ArrowLine(50,50)(0,10)
\ArrowLine(35,90)(50,50)
\ArrowLine(35,90)(80,90)
\DashLine(50,50)(90,50){2}
\put(-20,48){$(e)$}
\end{picture} \\

\caption{\label{fig1} \it The basic diagrams contributing in QCD to
  the formation of leptoquarks/squarks in $ep$ collisions. (a) Born
  term; (b) vertex loop; (c) gluon radiation off the
  leptoquark/squark; (d) gluon radiation off the quark; (e) gluon
  splitting to quarks and antiquarks. The same diagrams (a) to (d),
  interpreted backwards, describe the decay of leptoquarks/squarks to
  order $\alpha_s$.}
\end{figure}

In addition to the standard mass corrections for colored quarks and
bosons which are defined on the mass shells, the QCD corrections
involve the renormalization of the $e-q-\LQ/\squ$ vertex displayed in
Fig.~\ref{fig1}b. The $e-q-\LQ/\squ$ coupling $\lambda$ has been
defined at the point $\mu_R=M$ in the $\overline{MS}$ renormalization
scheme. The QCD vertex corrections make the coupling run according to
the formula
\begin{equation}
\lambda^2(\mu_R)=
\frac{\lambda^2(M)}{1+\frac{\alpha_S}{\pi} \log\frac{\mu_R^2}{M^2}}
\end{equation}
The second class of diagrams, Fig.~\ref{fig1}c/d, describe the
radiation of gluons off the initial quark and final leptoquark/squark
state: $e+q \to g+\LQ/\squ$. The sum of the virtual and bremsstrahlung
corrections is ultraviolet and infrared finite. Finally, the formation
process initiated by lepton-gluon collisions must be added: $e+g \to
\bar{q}+\LQ/\squ$, as shown in Fig.~\ref{fig1}e. The collinear
divergences due to gluon emission off the initial state quark line and
due to gluon splitting to quark-antiquark pairs will be absorbed in
the renormalization of the quark and gluon parton densities
\cite{alt2}.\bigskip

The final result of this simple analysis can be summarized in the
standard form
\begin{eqnarray}
\sigma(ep \rightarrow \LQ/\squ+X) & = & \sigma_{0} 
\left[ 1- \frac{2\pi^2}{9}\frac{\alpha_{s}}{\pi} \right]xq(x,M^2) +
\Delta \sigma_q + \Delta \sigma_g
\label{equ5}
\end{eqnarray}
with
\begin{eqnarray*}
\Delta \sigma_q & = & \int_{x}^{1} dx'~q(x',M^2)
\times \frac{\alpha_{s}}{\pi} \sigma_{0} \left\{ -\frac{z}{2}
P_{qq}(z) \log z + \frac{4}{3} [1 + z] \right. \nonumber \\
& & \hspace{12mm} \left. + \frac{4}{3} 
\left[ 2 \left(\frac{\log (1-z)}{1-z}
\right)_+ - \left( \frac{1}{1-z} \right)_+
- (2+z+z^2) \log (1-z) \right] \right\} \nonumber \\
\Delta \sigma_g & = & \int_{x}^{1} dx'~g(x',M^2)
\times \frac{\alpha_{s}}{\pi} \sigma_{0} \frac{z}{2} \left\{ - P_{qg}(z)
\left[\log \frac{z}{(1-z)^2} + 1\right] + z(1-z) \log z + 1 \right\}
\end{eqnarray*}
The auxiliary variable $z$ is defined as usual by $z=x/x'$. $P_{qq}$
and $P_{qg}$ are the standard $q \rightarrow q(g)$ and $g \rightarrow
q(\bar{q})$ splitting functions
\begin{eqnarray}
P_{qq}(z) & = & \frac{4}{3}\left\{ 2 \left( \frac{1}{1-z} \right)_+
-1-z+ \frac{3}{2}\delta(1-z)
\right\} \\
P_{qg}(z) & = & \frac{1}{2}\left\{z^2+(1-z)^2 \right\}
\end{eqnarray}
The distribution $F_+$ is defined by $F_+(z)=F(z)-\delta(1-z) \int_0^1
dz' F(z')$, regularizing the singularities at $z=1$. The
renormalization point $\mu_R$ and the factorization point $\mu_F$ have
been identified with the mass $M$ of the resonance. They may be
re-introduced as free parameters in Eq.~(\ref{eq5}) by substituting
$\lambda(M) \to \lambda(\mu_R)$ and adding the term $+\log
\mu_R^2/M^2$ to the coefficient of $\alpha_S/\pi$ in the virtual
correction; in addition, the mass scale $M$ in the parton densities
must be replaced by $\mu_F$ and $\log \mu_F/M^2$ must be added to the
logarithms attached to the parton splitting functions. \bigskip

\begin{figure}[hbt]
\vspace*{0.5cm}
\hspace*{0.8cm}
\begin{turn}{-90}%
\epsfxsize=9cm \epsfbox{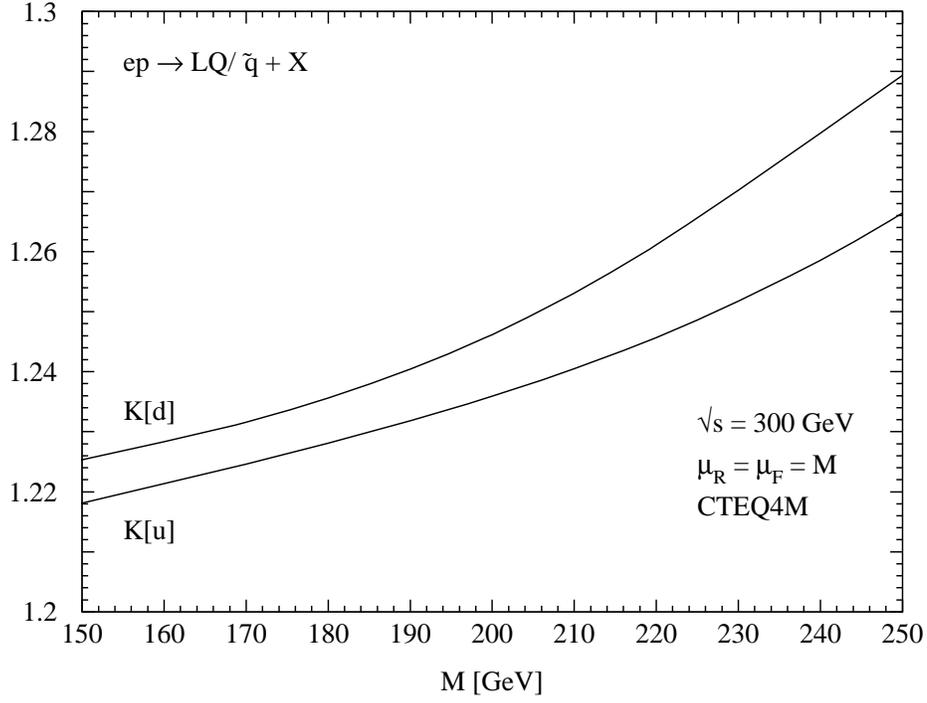}
\end{turn}
\vspace*{0.5cm}
\caption[]{\label{fig2} \it
  $K$-factors for $ed, eu \to \LQ/\tilde{q}$ as a function of the
  leptoquark/squark mass.}
\end{figure}

The results are exemplified in Fig.~\ref{fig2} for the $K$-factors
properly defined by
\begin{equation}
K = \sigma_{NLO}/\sigma_{LO}
\end{equation}
where all quantities (couplings, parton densities) in the numerator
and denominator are to be evaluated consistently in next-to-leading
and leading order, respectively. This definition guarantees the
correct theoretical interpretation of the size of the QCD corrections.
The $K$-factors are presented for $e^\pm$ projectiles on valence
($u,d$) targets in Fig.~\ref{fig2}. The $K$-factors are larger than
unity in the range of leptoquark masses we are interested in. The NLO
cross sections are larger than the leading order cross sections by
about 20 to 30\,\%.  The vertex corrections reduce the cross section,
as apparent from Eq.\ (\ref{equ5}). This effect is over-compensated
however by the quark contribution $\Delta \sigma_q$, while gluon
splitting, $\Delta \sigma_g$, plays a minor r\^ole. For sea quark
targets, the $K$-factor is close to 1.15 for $M = 150$ GeV and falls
to values below 1 for masses above 195 GeV. The different behavior of
$K$-factors for valence and sea quarks is a straight consequence of
the different shapes of the parton densities.
  
\begin{figure}[hbt]
\vspace*{0.5cm}
\hspace*{0.8cm}
\begin{turn}{-90}%
\epsfxsize=9cm \epsfbox{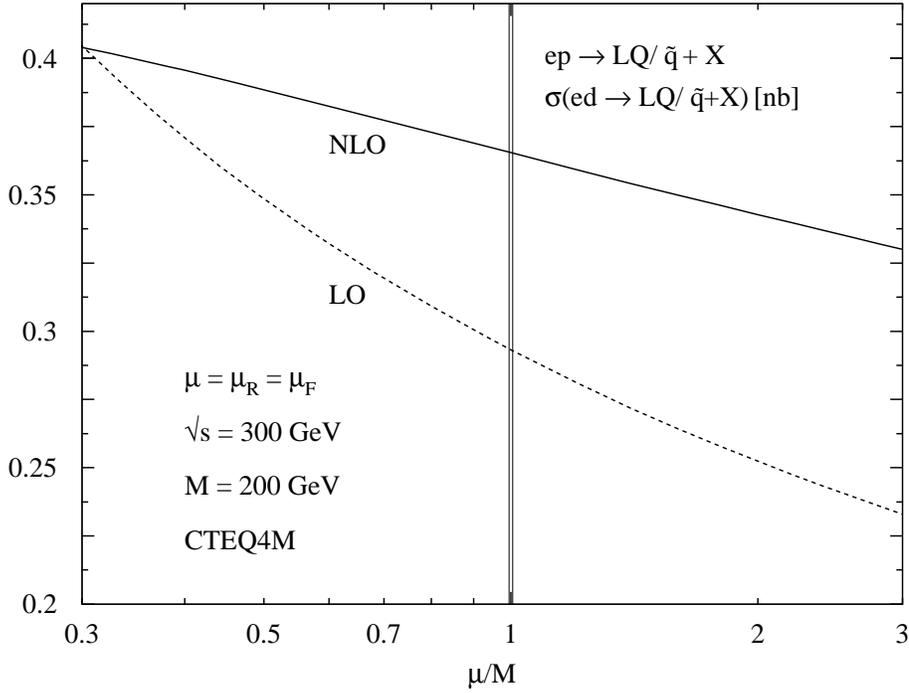}
\end{turn}
\vspace*{0.0cm}
\caption[]{\label{fig3} \it Comparison of the renormalization and
  factorization scale dependence in LO and NLO for the cross section
  for $\sigma(e + d \to \LQ/\tilde{q})$.}
\end{figure}

The improvement of the theoretical prediction is clearly visible in
Fig.~\ref{fig3}. The cross sections in next-to-leading order are
compared with the leading order for the process $e+d\to\LQ/\tilde{q}$
as a function of the renormalization scale $\mu_R$ and the
factorization scale $\mu_F$. $\mu_R$ and $\mu_F$ are identified for
the sake of simplicity. While the LO cross section changes by nearly a
factor of 2 if $\mu/M$ is varied from 3 to 1/3, the change of the NLO
cross section is damped to 1.2 in this range. Thus, if the
next-to-leading order corrections are taken into account, the
theoretical prediction of the cross section for scalar resonance
formation is stabilized considerably.\bigskip

\section{Decay Width and Lepton Spectrum}
The {\it partial width} for the decays 
\begin{equation}
\LQ/\squ \rightarrow e + q\, (g)
\end{equation}
can easily be calculated in a similar way \cite{sq}, including the
one-loop QCD corrections:
\begin{equation}
  \Gamma[\LQ/\squ \to eq(g)] = \frac{\lambda^2 M}{16 \pi} \left[ 1 +
  \left( \frac{9}{2} - \frac{4 \pi^2}{9} \right) \frac{\alpha_s}{\pi}
  \right]
\end{equation}
The coefficient in front of $\alpha_S/\pi$ is tiny, $\sim$~0.11, so
that QCD corrections to the width are small. For small Yukawa
couplings $\lambda \sim e/10$ the partial width is therefore very
narrow, i.e. $\Gamma \sim 3$~MeV for $M \sim 200$~GeV. If for
leptoquarks/squarks no other channels are open, apart from $\LQ \to e
q$ or $\nu q$, and $\squ \to eq$, the lifetime is very long, and the
leptoquark/squark will be bound into a fermionic ($\LQ \bar{q}$) or
($\squ\bar{q}$) state. This is different for squarks if the
branching ratio for $R$-parity violating decays is small compared to
$R$-parity conserving (neutralino/gaugino) decays.

To lowest order, the scaled lepton energy of the decay process
(\ref{decay}) in the rest-frame\footnote{The scaled lepton energy can
  be defined in a Lorentz-invariant way as $x_e =
  2(p_ep_{LQ/\tilde{q}})/M^2$.} of the isolated leptoquark or squark,
\begin{equation}
x_e = 2E_e / M ~~~~{\rm with}~~~~0 \leq x_e \leq 1
\end{equation}
is fixed to 1 if the final-state quark is treated as a massless
parton. However, perturbative gluon radiation (and the
non-perturbative hadronization of quark jets) reduces the lepton
energy and gives rise to a continuous energy spectrum. This effect is
analogous to the radiation of gluons in $q\bar{q}$ pair production of
$e^+e^-$ annihilation; LEP measurements can therefore serve as a solid
basis for estimating non-perturbative parameters which are not
accessible to theoretical QCD analyses.

\begin{figure}[hbt]
\vspace*{0.5cm}
\hspace*{0.8cm}
\epsfxsize=11cm \epsfbox{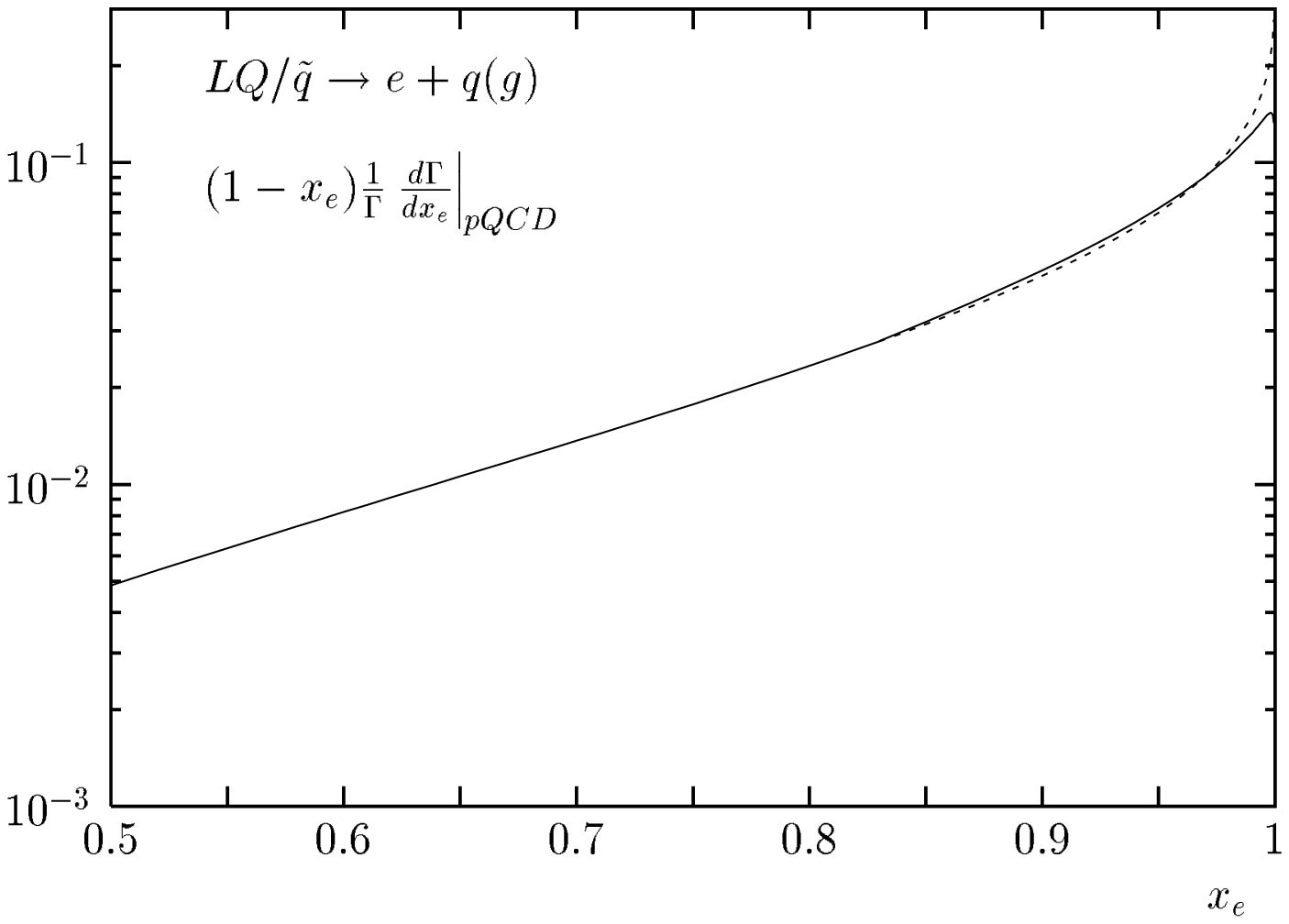}
\vspace*{0.0cm}
\caption[]{\label{fig4} \it The lepton-energy spectrum as predicted in
  perturbative QCD (full line). (The broken line corresponds to the
  ${\cal O}(\alpha_s)$ result without resummation near $x_e
  \rightarrow 1$).}
\end{figure}

The energy spectrum of the lepton, after real gluon emission is
switched on, is given by the expression
\begin{equation}
x_e < 1: ~~~~~~~~~~
\left.\frac{1}{\Gamma}\frac{d\Gamma}{dx_e}\right|^{pQCD}_{{\cal
      O}(\alpha_s)} =  
\frac{4}{3}\frac{\alpha_s}{\pi} 
\left[\frac{1}{x_1}\left(\log\frac{1}{x_1}-\frac{7}{4}\right)
+\log\frac{1}{x_1} + \frac{7}{4} - \frac{x_e}{4} \right]
\label{dgamma}
\end{equation}
with the abbreviation $x_1 = 1-x_e$.  This form of the spectrum is
normalized to the partial $eq$ decay width. (If other decay channels
are open, Eq.\ (\ref{dgamma}) must be supplemented by the branching
ratio $B_{eq}$ for absolute normalization). In the limit $x_e
\rightarrow 1$, the leading singularities may be resummed to give
\begin{equation}
x_e \rightarrow 1: ~~~~
\left.\frac{1}{\Gamma}\frac{d\Gamma}{dx_e}\right|^{pQCD}_{resum} =  
\left.\frac{1}{\Gamma}\frac{d\Gamma}{dx_e}\right|^{pQCD}_{{\cal
      O}(\alpha_s)}  
\times \exp\left(\frac{4}{3}\frac{\alpha_s}{\pi}\left[
-\frac{1}{2}\log^2\frac{1}{x_1}
-\frac{7}{4}\log\frac{1}{x_1}\right]\right)
\label{dgammar}
\end{equation}
The singularity of Eq.\ (\ref{dgamma}) for $x_e \rightarrow 1$ is
removed by resummation \cite{schier}. In fact, the spectrum
(\ref{dgammar}) approaches zero for $x_e \rightarrow 1$ and a sharp
maximum develops just below 1.  The final lepton-energy spectrum as
predicted in pQCD is shown by the full line in Fig.\ \ref{fig4}. The
spectrum in the approximation of one-gluon radiation without
resummation (singular near $x_e \rightarrow 1$) is indicated by the
broken line.

The softening of the lepton spectrum due to the perturbative gluon
radiation can be characterized by the average lepton energy $\langle
x_e \rangle$ defined for $\LQ \rightarrow e + q(g)$ decays. It is easy
to derive from the spectrum:
\begin{equation}
\langle x_e \rangle{}^{pQCD} = 
1 - \frac{4}{9} \frac{\alpha_s}{\pi}
\end{equation}
This corresponds to a loss of $\sim 3.4$ GeV in the lepton energy due
to perturbative gluon radiation for a leptoquark mass of 200 GeV.

In addition to the perturbative energy loss, residual non-perturbative
hadronization effects reduce the lepton energy, affecting the shape as
well as the average energy. Such effects can be parametrized in a
power series $\delta x_e^{NP} = -{\cal C}_1/M + \cdots$.  Estimates of
the size of the non-perturbative coefficient ${\cal C}_1$ may be
derived from the shape variable thrust \cite{abreu}. A closer analogon
is the direct-photon spectrum in $e^+e^-$ annihilation to hadrons
\cite{mattig} at LEP. With ${\cal C}_1 \sim 0.7$ GeV for thrust, the
non-perturbative contribution $\delta x_e^{NP} \sim$ ($-$ a few
$10^{-3}$) is expected well below the perturbative attenuation effect.
In Monte Carlo calculations, which are beyond the scope of the present
analysis, this qualitative estimate can be turned into a quantitative
prediction. The overall picture drawn from the pQCD calculation
presented above coincides well with the result of shower Monte Carlo
programs employed in the H1 analysis \cite{schleper}. This is expected
from the QCD analyses at LEP. Scenarios in which squarks decay so fast
that the color flux lines of the quarks in the decay final state are
connected directly to the proton remnant, require more elaborate Monte
Carlo programs.

\bigskip
\noindent {\large \bf Acknowledgements}\\[2ex] We are very grateful to
P.~M\"attig and P.~Schleper for discussions on experimental aspects.

 \newcommand{\zp}[3]{{\sl Z. Phys.} {\bf #1} (19#2) #3}
 \newcommand{\np}[3]{{\sl Nucl. Phys.} {\bf #1} (19#2)~#3}
 \newcommand{\pl}[3]{{\sl Phys. Lett.} {\bf #1} (19#2) #3}
 \newcommand{\pr}[3]{{\sl Phys. Rev.} {\bf #1} (19#2) #3}
 \newcommand{\prl}[3]{{\sl Phys. Rev. Lett.} {\bf #1} (19#2) #3}
 \newcommand{\prep}[3]{{\sl Phys. Rep.} {\bf #1} (19#2) #3}
 \newcommand{\fp}[3]{{\sl Fortschr. Phys.} {\bf #1} (19#2) #3}
 \newcommand{\nc}[3]{{\sl Nuovo Cimento} {\bf #1} (19#2) #3}
 \newcommand{\ijmp}[3]{{\sl Int. J. Mod. Phys.} {\bf #1} (19#2) #3}
 \newcommand{\ptp}[3]{{\sl Prog. Theo. Phys.} {\bf #1} (19#2) #3}
 \newcommand{\sjnp}[3]{{\sl Sov. J. Nucl. Phys.} {\bf #1} (19#2) #3}
 \newcommand{\cpc}[3]{{\sl Comp. Phys. Commun.} {\bf #1} (19#2) #3}
 \newcommand{\mpl}[3]{{\sl Mod. Phys. Lett.} {\bf #1} (19#2) #3}
 \newcommand{\cmp}[3]{{\sl Commun. Math. Phys.} {\bf #1} (19#2) #3}
 \newcommand{\jmp}[3]{{\sl J. Math. Phys.} {\bf #1} (19#2) #3}
 \newcommand{\nim}[3]{{\sl Nucl. Instr. Meth.} {\bf #1} (19#2) #3}
 \newcommand{\el}[3]{{\sl Europhysics Letters} {\bf #1} (19#2) #3}
 \newcommand{\ap}[3]{{\sl Ann. of Phys.} {\bf #1} (19#2) #3}
 \newcommand{\jetp}[3]{{\sl JETP} {\bf #1} (19#2) #3}
 \newcommand{\jetpl}[3]{{\sl JETP Lett.} {\bf #1} (19#2) #3}
 \newcommand{\acpp}[3]{{\sl Acta Physica Polonica} {\bf #1} (19#2) #3}
 \newcommand{\vj}[4]{{\sl #1~}{\bf #2} (19#3) #4}
 \newcommand{\ej}[3]{{\bf #1} (19#2) #3}
 \newcommand{\vjs}[2]{{\sl #1~}{\bf #2}}
 \newcommand{\hep}[1]{{\sl hep--ph/}{#1}}
 \newcommand{\desy}[1]{{\sl DESY-Report~}{#1}}

\newpage

\begin{thebibliography}{99}
\bibitem{H1}    C.~Adloff et al., H1 Collab., DESY-97-024, {\it
    Z.~Phys.} {\bf C} in press.
\bibitem{ZEUS}  J.~Breitweg et al., Zeus Collab., DESY-97-025, {\it
    Z.~Phys.} {\bf C} in press.
\bibitem{LEP}   S.\ Komamiya, for the OPAL collaboration, CERN Seminar,
                Feb.~25th, 1997;
                P.\ M\"attig, Proceedings, {\it International
                  Conference on High Energy Physics} (Warsaw 1996);
                OPAL Physics Note PN 280,
                and P.\ M\"attig, private communication.
\bibitem{TEV}   D0 collaboration, 
{\tt http://d0wop.fnal.gov/public/new/lq/lq\_\_blurb.html}. 
\bibitem{alta}  G.~Altarelli, J.~Ellis, G.F.~Giudice, S.~Lola, 
                and M.L.~Mangano, CERN-TH-97-040 (hep-ph/9703276).
\bibitem{kali}  J.~Kalinowski, R.~R\"uckl, H.~Spiesberger,
                and P.M.~Zerwas, DESY 97-38 (hep-ph/9703288).
\bibitem{blum}  J.~Bl\"umlein, DESY-97-32 (hep-ph/9703287).
\bibitem{wilc}  K.S.~Babu, C.~Kolda, J.~March-Russell, and F.~Wilczek,
                IASSNS-HEP-97-04 (hep-ph/9703299).
\bibitem{bhat}  D.~Choudhury and S.~Raychaudhuri, CERN-TH/97-26
    (hep-ph/9702392); D.~Choudhury and S.~Raychaudhuri, CERN-TH/97-51
    (hep-ph/9703369);   
\bibitem{drei}  H.~Dreiner and P.~Morawitz, RAL report (hep-ph/9703287). 
\bibitem{buch}  J.~Wudka, \pl{167B}{86}{337};  
                W.~Buchm\"uller, R.~R\"uckl, and D.~Wyler,
                \pl{B191}{87}{442}.
\bibitem{alt2}  G.~Altarelli, R.K.~Ellis, and G.~Martinelli,
                \np{B157}{79}{461};
                W.~Furmanski and R.~Petronzio,
                \zp{C11}{82}{293}.
\bibitem{sq}    W.~Beenakker, R.~H\"opker, and P.M.~Zerwas,
                \pl{B378}{96}{159}.
\bibitem{schier} G.~Schierholz, SLAC Summer Inst.\ 1979, p.\ 476;
                P.~Bin\'etruy, {\it Phys.\ Lett.}\ {\bf 91B} (1980) 245. 
\bibitem{abreu} P.~Abreu et al., Delphi Collab., \zp{C73}{97}{229}.
\bibitem{mattig} R.~Akers et al., Opal Collab., \zp{C67}{95}{15}.
\bibitem{schleper} P.~Schleper, private communication.
\end{thebibliography}
\end{document}